\documentclass{article}
\usepackage{spconf,amsmath,graphicx}
\usepackage{amssymb}
\usepackage{enumitem}
\setlist[itemize]{leftmargin=*, nosep}
\usepackage{float}
\usepackage{booktabs}
\usepackage{microtype}
\usepackage[table]{xcolor}   % 必须在 arydshln 之前
\usepackage{arydshln}        % 必须在 [table]xcolor 之后
\usepackage{tabularx}
\usepackage{multirow}
\usepackage{makecell}
\usepackage{soul}
\usepackage{hyperref}        % 最后加载
\definecolor{myblue}{rgb}{0.933, 0.933, 0.996}
\definecolor{myred}{rgb}{0.996, 0.925, 0.925} 
\definecolor{mygray}{rgb}{0.914, 0.914, 0.914}
\definecolor{mypurple}{rgb}{0.945, 0.925, 0.976}
\definecolor{lightpurple}{RGB}{220, 200, 255}
\sethlcolor{lightpurple}
\newcolumntype{W}{>{\centering\arraybackslash}X}
\newcolumntype{Z}{>{\raggedleft\arraybackslash}X}
\newcolumntype{Y}{>{\centering\arraybackslash}X}

\title{WaveScat: Wavelet Scattering Front-Ends with Self-Supervised Features for Speech Deepfake Detection}
\name{\thanks{The code is available at https://github.com/xxuan-acoustics/WaveScat}.}
\name{Xi Xuan$^{1,2}$, Davide Carbone$^{3}$, Wenxin Zhang$^{4}$, Tomi H. Kinnunen$^{1}$}
\address{$^{1}$School of Computing, University of Eastern Finland, Finland\\
         $^{2}$Beckman
Institute for Advanced Science and Technology, University of Illinois Urbana-Champaign, USA\\
         $^{3}$Laboratoire de Physique de l'\'Ecole Normale Sup\'erieure, ENS, Universit\'e PSL, CNRS,\\
         \phantom{$^{3}$}Sorbonne Universit\'e, Universit\'e de Paris, France\\
         $^{4}$School of Computer Science and Technology, University of Chinese Academy of Sciences, China}

\begin{document}
\ninept

\maketitle

\vspace{-0.5cm}
\begin{abstract}
Existing front-ends for speech deepfake detection are primarily categorized into two types. Hand-crafted filterbank features are transparent but limited in capturing higher-level information. SSL features, in turn, lack interpretability and may overlook fine-grained spectral anomalies. We propose WaveScat, a novel family of feature extractors that combines the best of both worlds via the wavelet scattering transform (WST), which cascades wavelet convolutions with modulus nonlinearities to produce deformation-stable, multi-scale features. Experiments on the recent Deepfake-Eval-2024 benchmark, together with cross-dataset evaluations on SpoofCeleb, In-the-Wild, and ASVspoof 5, show that WaveScat outperforms existing front-ends by a wide margin. Our analysis reveals that a small averaging scale combined with high-frequency and directional resolutions is critical for capturing subtle artifacts. This underscores the value of stable and translation-invariant features for speech deepfake detection. The code and supplementary materials are available online \footnote{https://github.com/xxuan-acoustics/WaveScat}.
\end{abstract}

\section{Introduction}
\label{sec:intro}

Speech deepfake detectors (SDDs) aim to distinguish artificially generated speech from real human speech. 
SDD systems consist of a front-end (feature extractor) \cite{xuan2025wavesp,ye2025amplifying, guo2024audio} followed by a back-end (classifier) \cite{xuan2025fakemamba,tran25b_interspeech,gu2025allm4add}. The choice of the former is critical as it determines the SDD's ability to capture subtle acoustic artifacts suitable for deepfake detection. SDD front-ends can be broadly categorized into digital signal processing (DSP) and self-supervised learning (SSL) approaches, each offering distinct advantages. 

The former category utilizes time-frequency analysis techniques based on the short-time Fourier transform (STFT) \cite{sahidullah2015comparison}, including mel-spectrograms \cite{fathan2022mel} 
and linear-frequency cepstral coefficients (LFCCs)  \cite{sahidullah2015comparison}. 
These representations are obtained by applying a bank of frequency-localized filters to the STFT magnitude spectrum, resulting in a low-dimensional representation. Early studies also used the constant-Q transform, a multiresolution time-frequency method adopted from music signal processing, to extract corresponding cepstral coefficients, CQCCs \cite{TODISCO2017516}. Despite transparency, simplicity, and computational efficiency, these features are limited in their lack of robustness, suboptimal time-frequency analysis properties, and spectral smoothing introduced by the filterbank. 

On the other hand, modern SSL models, such as XLSR \cite{ref19}, HuBERT \cite{hsu2021hubert}, and MMS \cite{pratap2024scaling}, provide a robust, data-driven alternative to hand-crafted filterbank feature extraction. They are trained on massive amounts of data, leveraging data augmentation and masking techniques to learn representations robust to noisy or missing observations \cite{hsu2021hubert}. While SSLs outperform pure DSP front-ends in detection and generalization, they suffer from high computational costs and limited explainability.
This limitation is a particular concern in audio forensic investigations, where feature interpretability is not merely a desirable property but a fundamental requirement; scientific evidence should be transparent, reproducible, and open to scrutiny \cite{iso30107}.

To address these shortcomings, our work introduces, for the first time, the \textit{wavelet scattering transform} (WST)  \cite{mallat2012group,bruna2013invariant, valogiannis2022towards} to SDD. WST serves as a bridge between DSP-based and data-driven front-ends, as it can be interpreted as a mathematical counterpart to convolutional layers in neural networks. \emph{Importantly, WST requires no training data} but is entirely defined through invariance and stability properties concerning signal translation and deformation. Furthermore, the hierarchical structure of the scattering coefficients offers a clear physical interpretation of multiscale processes, making it suitable for analyzing speech and sound signals  \cite{khatami2018origins,licciardi2024whalenet}.

We introduce WaveScat, a novel family of wavelet-scattering 
front-ends for SDD that combines the mathematical transparency of wavelets with the representational power of SSL features. In particular, we propose two WaveScat feature extractor designs, as illustrated in Fig.~\ref{fig:3} and detailed in Section~\ref{subsec:wstx_extractor}. A 
systematic analysis of their control parameters, together with the scattering 
energy distribution, reveals which configurations best expose synthesis 
artifacts. Experiments on Deepfake-Eval-2024 \cite{chandra2025deepfakeeval2024} and 
three cross-dataset benchmarks (SpoofCeleb \cite{jung2025spoofceleb}, 
In-the-Wild \cite{muller22_interspeech}, ASVspoof~5 \cite{wang2026asvspoof}) 
confirm both generalization and interpretability. The latter is substantiated by SHAP feature importance analysis \cite{NIPS2017_8a20a862} on the scattering coefficients. WaveScat attains a low real-time factor (RTF) with a parameter budget comparable to existing front-ends, demonstrating its efficiency for real-time audio forensics.

% \vspace{-0.5cm}
\section{Proposed Method}
% \vspace{-0.2cm}
\subsection{Wavelet Scattering Transform Theory}
\begin{figure}[H]
    \centering
    \vspace{-0.3cm}
    \includegraphics[width=0.48\textwidth]{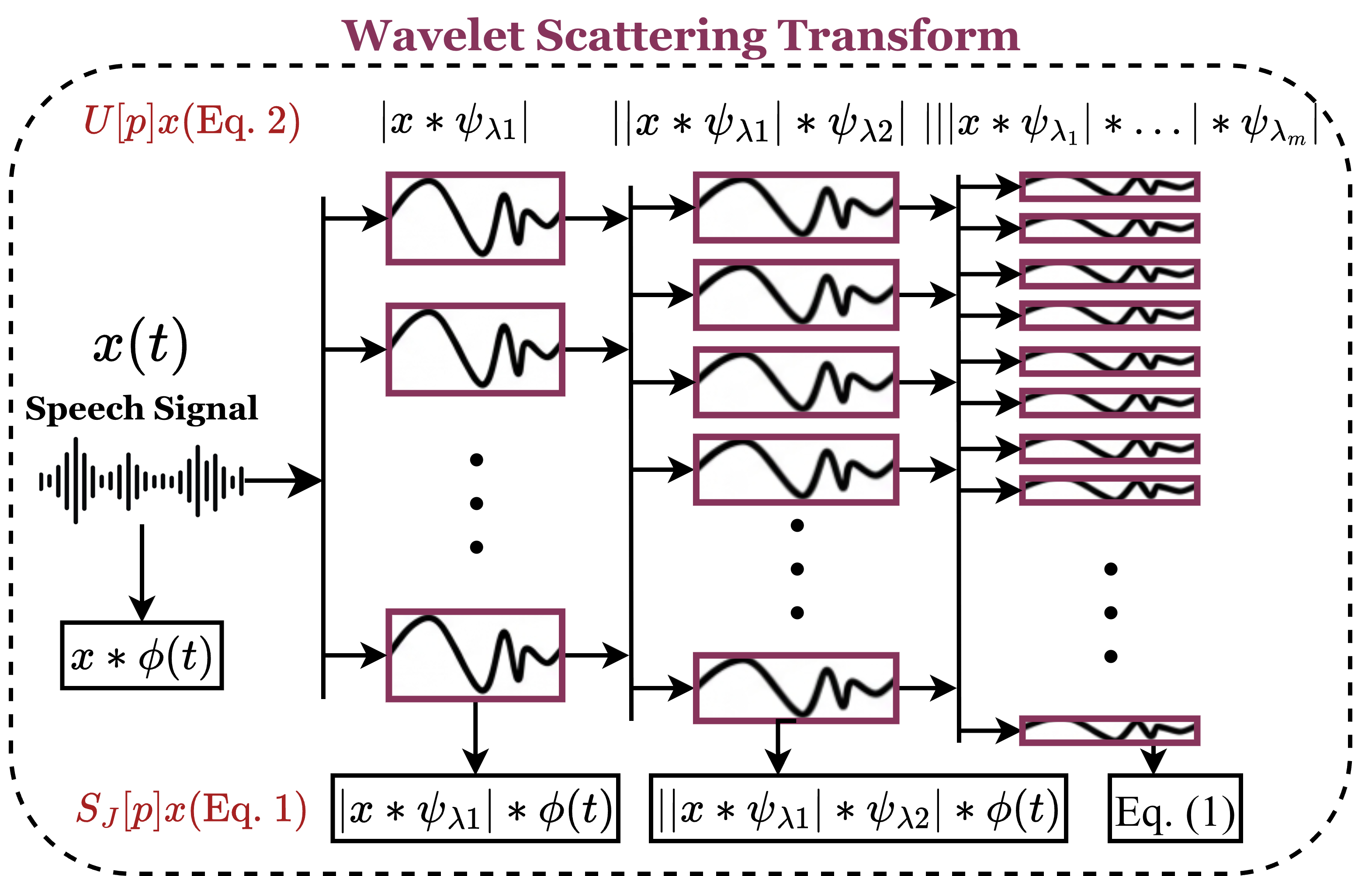}
    \caption{\footnotesize Hierarchical architecture of the second-order wavelet scattering transform, showing the extraction of zeroth-, first-, and second-order coefficients.}
    \label{fig:structure}
    \vspace{-0.3cm}
\end{figure}

The wavelet scattering transform (WST) \cite{mallat2012group,bruna2013invariant} stands as a mathematical operator capable of yielding a stable and translation-invariant representation for a speech signal $x(t)$ through a cascade of wavelet modulus operators, as illustrated in Fig. \ref{fig:structure}. 

For discrete signals sampled at $f_s$, this analysis is performed over a temporal invariance scale $T \!=\! 2^J / f_s$ seconds, where $2^J$ denotes the window size. This cascade is parameterized by a path $p \!= \!(\lambda_1, \dots, \lambda_m)$, defined as a tuple of length $m$ built using indices $\lambda_i \in \Lambda^J$ representing an ordered sequence of wavelet scales; the wavelet scattering coefficient $S_J[p]x(t)$ along $p$ is the convolution of a propagation operator $U[p]x$ with a scaled Gaussian low-pass filter $\phi_{2^J}$:
\begin{equation} 
S_J[p]x(t) = (U[p]x \ast \phi_{2^J})(t) \!= \!\int_{-\infty}^{\infty} U[p]x(\tau) \phi_{2^J}(t-\tau) \, d\tau, \label{eq:wst_main} 
\end{equation}
where the nonlinear cascade operator is defined as:
\begin{equation}
U[\lambda]x(t)=\big|(x \ast \psi_\lambda)(t)\big|, \quad
U[p]x = U[\lambda_m]\cdots U[\lambda_1]x .
\label{eq:cascade}
\end{equation}
Here, $\ast$ denotes convolution and $\phi_{2^J}(t)\!=\!2^{-J}\phi(t/2^J)$. We define wavelets as $\psi_\lambda(t)\!=\!\lambda^{-1}\psi(t/\lambda)$, which
preserves the $L^1$-norm across scales, following Kymatio~\cite{andreux2020kymatio},
an open-source library for scattering transforms. The continuous $t$ in
Eqs.~(1)--(2) is discretized in the implementation.

To effectively capture the spectral richness of speech signals, we employ 
scales $\lambda \in \Lambda^J = \{2^{j/Q}\}_{j=0}^{JQ-1}$, where $Q$ is the number 
of wavelets per octave that determines the log-frequency sampling resolution. These range from dyadic scales (i.e., powers of two) to finer intermediate scales, restricted to remain finer than the averaging scale $2^J$. While wavelet transforms  \cite{beylkin1991fast} provide stability under the action of small diffeomorphisms, the nonlinear operation and the integration over time yield translation invariance  \cite{mallat2012group}. Higher-order cascades recover high-frequency amplitude modulations of the first-order envelopes that are lost due to the low-pass averaging in lower-order coefficients.
Furthermore, the WST is robust to additive noise \cite{licciardi2024whalenet}, a property guaranteed by the cascaded wavelet-modulus 
structure \cite{wiatowski2017mathematical}. The theoretical noise robustness has been validated in pathological voice 
recognition \cite{souli2021robust} and underwater acoustic 
classification \cite{liu2025underwater}. 

% \vspace{-0.5cm}
\subsection{Physical Interpretation of WST Parameters}
\label{subsec:wst_params}
\vspace{-0.2cm}
We use the Kymatio library~ \cite{andreux2020kymatio} to implement both 1D and 2D WSTs, which forms the foundational basis for our front-ends. The 1D transform operates directly on raw waveforms, whereas the 2D transform processes time-frequency representations, such as spectrograms. The theoretical foundation established above pertains to 1D signals, whereas an extension to higher dimensions (2D WST) can be found in~ \cite{bruna2013invariant}. Complex Morlet wavelets~ \cite{stephane1999wavelet} are utilized for the cascaded operations. The 1D WST is characterized by three primary control parameters. The \emph{averaging scale} $J$ ($J\! \ge\! 2$) determines the window size $2^J$, with smaller $ J$ preserving higher-frequency temporal details. The \emph{number of wavelets per octave} $Q$ ($Q \!\ge\! 1$) determines the frequency resolution. Finally, the \emph{scattering order} $M$ sets the maximum path length, 
accumulating features from zeroth-order time averages, first-order
spectral envelopes, and second-order amplitude-modulation 
coefficients \cite{anden2014deep}.

For the 2D WST, we use SSL latent feature maps as input, viewed as two-dimensional images in the (time, feature) plane, analogous to time-frequency representations but with spectral magnitudes replaced by SSL features. The 2D WST is characterized by three hyperparameters. The \emph{averaging scale} $J$, which defines the maximum spatial scale of the 2D 
low-pass filter in powers of 2, governs the degree of averaging across time and 
feature axes. The \emph{angular resolution} $L$ represents the number of 
orientations in the 2D wavelet bank to provide directional selectivity. The 
\emph{scattering order} $M$ is defined as in the 1D case. Because SSL latent features lack time or frequency axes, these parameters cannot be directly interpreted in terms of time or frequency variables. However, the 2D wavelet representation captures the multi-scale structure inherent in the SSL latent features, leading to improved detection performance (demonstrated below).

% \vspace{-0.5cm}
\subsection{WaveScat Feature Extractor}
\begin{figure}[H]
    \centering
    \scriptsize
    \includegraphics[width=0.45\textwidth]{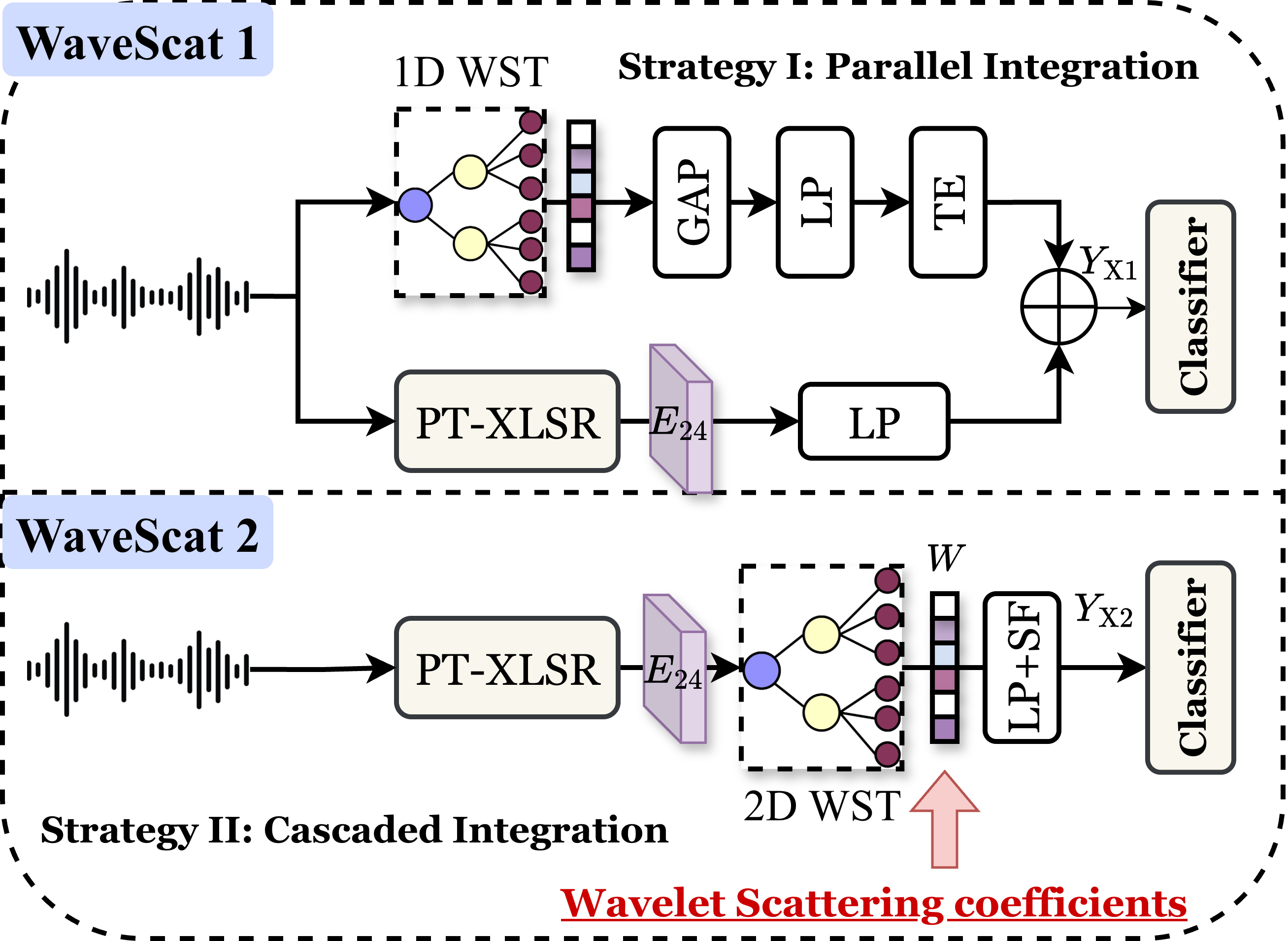}
    % \vspace{-0.3cm}
    \caption{\footnotesize Overview of the WaveScat series: WaveScat1 and WaveScat2 feature extractors. The top panel illustrates Strategy I (parallel integration with 1D WST), while the bottom panel shows Strategy II (cascaded integration with 2D WST). GAP (Global Average Pooling); LP (Linear Projection); TE (Temporal Expansion); SF (Spatial Flattening).}
    \label{fig:3}
    \vspace{-0.4cm}
\end{figure}

\vspace{-0.2cm}
\label{subsec:wstx_extractor}
In principle, WST could be used as a standalone acoustic front-end similar to MFCCs, LFCCs, or CQCCs~ \cite{sahidullah2015comparison, TODISCO2017516}. However, by capitalizing on the robustness of SSL models such as XLSR~ \cite{xin2022investigating,xuan25_spsc}, we propose WaveScat, a series of front-ends that combine WST and SSL features to bridge the gap between physical signal representations and SSL embeddings. As shown in Fig.~\ref{fig:3}, WaveScat comprises two architectural designs: \emph{parallel} (WaveScat1) and \emph{cascaded} (WaveScat2), detailed as follows.

\vspace{0.1cm}
\noindent{\textbf{Prompt Tuning XLSR (PT-XLSR).}} We adopt XLSR-300M  \cite{ref19} as the foundation model, adhering to the parameter-efficient prompt tuning setup (PT-XLSR) described in  \cite{xuan2025wavesp, xie2025detect}. We freeze the XLSR parameters and introduce $k$ learnable prompt tokens $V_i \!\in \!\mathbb{R}^{k \times D}$ at each transformer layer $i \in \{1,\ldots,24\}$, where $D$ denotes the hidden dimensionality. The CNN-extracted features \mbox{$E_0\!\in \!\mathbb{R}^{N \times D}$}, where $N$ is the number of time frames, are concatenated with the prompt tokens to guide the encoding process, producing the final representation \mbox{$E_{24}\! \in\! \mathbb{R}^{(k+N) \times D}$} for subsequent integration strategies.

\vspace{0.1cm}
\noindent{\textbf{Strategy I: Parallel Integration (WaveScat1).}} {WaveScat1} is formulated as a parallel dual-branch architecture comprising the {1D WST} and {PT-XLSR} components, both of which operate directly on the raw waveform. To align the feature spaces from both branches, the {1D WST} branch extracts scattering coefficients and processes them via global average pooling, linear projection, and temporal expansion. Concurrently, the {PT-XLSR} branch linearly projects {$E_{24}$} from the transformer hidden dimension {$D$} to 144. Finally, the outputs from both branches are concatenated channel-wise, resulting in the fused representation \mbox{$Y_{\text{X1}} \in \mathbb{R}^{(k+N) \times 288}$}.

\begin{table}[t]
\centering
\scriptsize
\setlength{\tabcolsep}{2pt}
\renewcommand{\arraystretch}{1.3}
\caption{\footnotesize Deepfake-Eval-2024 results for WaveScat feature extractor (FE) under different parameter settings, each combined with a shared Mamba-based classifier. Best results are in \textbf{bold}. Confidence intervals are in parentheses.}
\label{tab:1}
\resizebox{\columnwidth}{!}{%
\begin{tabular}{l | c c c | c c c c}
\noalign{\hrule height 1.2pt}
\textbf{FE} & \textbf{J} & \textbf{Q} & \textbf{M} & \textbf{minDCF}$\downarrow$ & \textbf{EER(\%)}$\downarrow$ & \textbf{F1(\%)}$\uparrow$ & \textbf{AUC(\%)}$\uparrow$ \\
\noalign{\hrule height 1.2pt}
\multirow{8}{*}{\rotatebox{90}{\textbf{WaveScat1}}}
 & 2 & 1 & 2 & \makecell{0.3540 \tiny ($\pm$0.0157)} & \makecell{15.19 \tiny ($\pm$0.52)} & \makecell{82.14 \tiny ($\pm$0.62)} & \makecell{92.13 \tiny ($\pm$0.37)} \\
 & 2 & 8 & 2 & \makecell{0.3682 \tiny ($\pm$0.0139)} & \makecell{14.98 \tiny ($\pm$0.50)} & \makecell{78.37 \tiny ($\pm$0.66)} & \makecell{89.45 \tiny ($\pm$0.58)} \\
 & \cellcolor{myred}2 & \cellcolor{myred}10 & \cellcolor{myred}2 & \cellcolor{myred}\makecell{\textbf{0.3408} \tiny ($\pm$0.0161)} & \cellcolor{myred}\makecell{\textbf{14.18} \tiny ($\pm$0.63)} & \cellcolor{myred}\makecell{\textbf{81.66} \tiny ($\pm$0.58)} & \cellcolor{myred}\makecell{\textbf{92.50} \tiny ($\pm$0.40)} \\
\cdashline{2-8}
 & 4 & 10 & 2 & \makecell{0.4182 \tiny ($\pm$0.0099)} & \makecell{15.04 \tiny ($\pm$0.41)} & \makecell{76.85 \tiny ($\pm$0.67)} & \makecell{90.35 \tiny ($\pm$0.47)} \\
 & 6 & 10 & 2 & \makecell{0.4172 \tiny ($\pm$0.0087)} & \makecell{17.20 \tiny ($\pm$0.89)} & \makecell{76.53 \tiny ($\pm$0.42)} & \makecell{90.84 \tiny ($\pm$0.24)} \\
 & 8 & 10 & 2 & \makecell{0.4782 \tiny ($\pm$0.0122)} & \makecell{16.77 \tiny ($\pm$0.41)} & \makecell{79.11 \tiny ($\pm$0.63)} & \makecell{89.93 \tiny ($\pm$0.34)} \\
\cdashline{2-8}
 & 2 & 10 & 3 & \makecell{0.4147 \tiny ($\pm$0.0127)} & \makecell{14.93 \tiny ($\pm$0.54)} & \makecell{80.78 \tiny ($\pm$0.69)} & \makecell{91.23 \tiny ($\pm$0.34)} \\
 & 2 & 10 & 1 & \makecell{0.3901 \tiny ($\pm$0.0126)} & \makecell{16.37 \tiny ($\pm$0.58)} & \makecell{75.40 \tiny ($\pm$0.70)} & \makecell{90.82 \tiny ($\pm$0.44)} \\
\noalign{\hrule height 1.2pt}
\textbf{FE} & \textbf{J} & \textbf{L} & \textbf{M} & \textbf{minDCF}$\downarrow$ & \textbf{EER(\%)}$\downarrow$ & \textbf{F1(\%)}$\uparrow$ & \textbf{AUC(\%)}$\uparrow$ \\
\noalign{\hrule height 1.2pt}
\multirow{9}{*}{\rotatebox{90}{\textbf{WaveScat2}}}
 & 2 & 1 & 2 & \makecell{0.4852 \tiny ($\pm$0.0167)} & \makecell{17.00 \tiny ($\pm$0.52)} & \makecell{75.19 \tiny ($\pm$0.83)} & \makecell{85.98 \tiny ($\pm$0.48)} \\
 & 2 & 8 & 2 & \makecell{0.3703 \tiny ($\pm$0.0142)} & \makecell{14.94 \tiny ($\pm$0.52)} & \makecell{79.88 \tiny ($\pm$0.60)} & \makecell{90.08 \tiny ($\pm$0.37)} \\
 & \cellcolor{myblue}2 & \cellcolor{myblue}10 & \cellcolor{myblue}2 & \cellcolor{myblue}\makecell{\textbf{0.3567} \tiny ($\pm$0.0081)} & \cellcolor{myblue}\makecell{\textbf{14.84} \tiny ($\pm$0.40)} & \cellcolor{myblue}\makecell{\textbf{81.83} \tiny ($\pm$0.47)} & \cellcolor{myblue}\makecell{\textbf{92.43} \tiny ($\pm$0.23)} \\
\cdashline{2-8}
 & 2 & 6 & 3 & \makecell{0.3811 \tiny ($\pm$0.0095)} & \makecell{14.96 \tiny ($\pm$0.42)} & \makecell{81.09 \tiny ($\pm$0.58)} & \makecell{90.16 \tiny ($\pm$0.33)} \\
 & 2 & 8 & 3 & \makecell{0.3883 \tiny ($\pm$0.0164)} & \makecell{15.24 \tiny ($\pm$0.74)} & \makecell{79.23 \tiny ($\pm$0.78)} & \makecell{91.18 \tiny ($\pm$0.46)} \\
 & 2 & 10 & 3 & \makecell{0.3743 \tiny ($\pm$0.0087)} & \makecell{14.98 \tiny ($\pm$0.21)} & \makecell{81.57 \tiny ($\pm$0.49)} & \makecell{91.40 \tiny ($\pm$0.36)} \\
\cdashline{2-8}
 & 3 & 8 & 1 & \makecell{0.4764 \tiny ($\pm$0.0141)} & \makecell{17.24 \tiny ($\pm$0.69)} & \makecell{78.71 \tiny ($\pm$0.69)} & \makecell{85.60 \tiny ($\pm$0.59)} \\
 & 3 & 8 & 2 & \makecell{0.4284 \tiny ($\pm$0.0153)} & \makecell{17.81 \tiny ($\pm$0.50)} & \makecell{78.94 \tiny ($\pm$0.59)} & \makecell{89.50 \tiny ($\pm$0.40)} \\
 & 3 & 8 & 3 & \makecell{0.4939 \tiny ($\pm$0.0146)} & \makecell{18.21 \tiny ($\pm$0.46)} & \makecell{76.48 \tiny ($\pm$0.74)} & \makecell{87.22 \tiny ($\pm$0.56)} \\
\noalign{\hrule height 1.2pt}
\end{tabular}%
}
\vspace{-0.5cm}
\end{table}

\vspace{0.1cm}
\noindent{\textbf{Strategy II: Cascaded Integration (WaveScat2).}} {WaveScat2} uses a cascaded single-pathway architecture. The waveform is first processed by {PT-XLSR} to extract high-level SSL latent feature maps {$E_{24}$}, which are fed into a {2D WST} to characterize intra-channel temporal dynamics and inter-channel structural correlations, obtaining a scattering tensor \mbox{$W \!\in \!\mathbb{R}^{C_{\mathrm{path}} \times N' \times D_{\mathrm{scat}}}$}. Here, \mbox{$N'\! =\! \lfloor (k+N)/2^J \rfloor$} and {$D_{\mathrm{scat}}\! =\! \lfloor D/2^J \rfloor$} denote the downsampled temporal and spectral resolutions, respectively. The number of scattering channels {$C_{\mathrm{path}}$} is determined by concatenating coefficients up to the second order  \cite{bruna2013invariant}, comprising 1 zeroth-order, {$JL$} first-order, and {$L^2 \cdot \binom{J}{2}$} second-order paths. Thus, the total dimension is {$C_{\mathrm{path}}\!=\!1\!+\!JL\!+\!L^2J(J\!-\!1)/2$}. These scattering coefficients capture multi-order spectro-temporal details. Finally, a linear projection reduces {$D_{\mathrm{scat}}$} to 144, followed by a spatial flatten to reshape the tensor into {$Y_{\mathrm{X2}} \!\in \!\mathbb{R}^{L_{\mathrm{seq}} \times 144}$}, where \mbox{$L_{\mathrm{seq}} \!= \!N' \times C_{\mathrm{path}}$}.

\noindent{\textbf{Classifier.}} The extracted feature {$\mathbf{Y} \in \{Y_{X1}, Y_{X2}\}$} is fed into a recent and robust classifier~ \cite{xuan2025fakemamba} to produce a probability score $\hat{p}$, thereby classifying the input speech as real or fake.

% \vspace{-0.4cm}
\section{Experimental setup}
\label{sec:Experiment}

\subsection{Datasets}
Our main experiments use Deepfake-Eval-2024 (DE2024) \cite{chandra2025deepfakeeval2024}, a benchmark of in-the-wild deepfakes comprising 56.5 hours of real and fake audio collected from social media in 2024, spanning 40 languages. Following~ \cite{xuan2025wavesp}, 
we sliced the official training and test sets into non-overlapping 
4-second chunks, resulting in $\sim$50k wav files, and split the 
training set into training and development sets at a 9:1 ratio. To further evaluate generalization, all models are trained on DE2024 and 
evaluated on three additional benchmarks as test sets: 
SpoofCeleb~\mbox{ \cite{jung2025spoofceleb}} ($\sim$91k utterances), 
In-the-Wild (ITW)~\mbox{ \cite{muller22_interspeech}} ($\sim$32k utterances), 
and ASVspoof~5~\mbox{ \cite{wang2026asvspoof}} ($\sim$681k utterances).

\begin{figure*}
    \centering
    \scriptsize
    \includegraphics[width=0.99\linewidth]
    {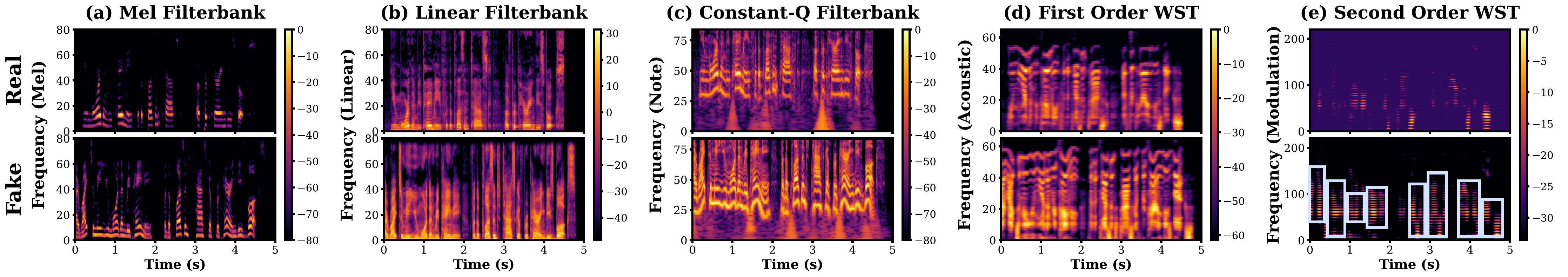}  
    %\scriptsize

    \caption{\footnotesize Representations of a real utterance (top row) and a fake utterance synthesized by Qwen2.5-Omni (bottom row) across different front-ends: (a) Mel, (b) Linear, (c) Constant-Q Filterbank, (d) First-order WST, and (e) Second-order WST. The displayed WST representations correspond to the configuration $(J, Q) = (2, 10)$. Focusing on the bottom row, the correspondence between larger WST scales and lower spectrogram frequencies is visually evident. Notably, the preceding three spectrogram representations appear more coarse-grained than the first- and second-order WST. The blue bounding boxes highlight the visually distinctive parts of the fake speech signals within the WST features compared to real speech.}    
    \vspace{-0.3cm}
    \label{fig:feature_comparison}
\end{figure*}
\vspace{-0.2cm}
\subsection{Model Configurations}
% \vspace{-0.2cm}
We evaluate SDD systems by pairing different front-ends with a shared classifier~ \cite{xuan2025fakemamba}. We use Librosa~ \cite{mcfee2015librosa} to downsample the raw audio to 16 kHz and 
extract the mel-scale, linear, and constant-Q (CQ) log filterbank outputs, 
without the subsequent discrete cosine transform as in cepstral feature 
computation. To ensure a fair comparison, all features are extracted using a 10 ms hop length. Mel and linear filterbank features employ a 25 ms frame size and a Hanning window, while CQ filterbank uses 9 bins per octave to define the frequency resolution, resulting in feature matrices of shape $(80, 399)$. 1D and 2D WST were implemented using Kymatio~ \cite{andreux2020kymatio}. To explore suitable control parameters (defined in Sec.~\ref{subsec:wst_params}), we conducted experiments by selecting $J$, $Q$, and $M$ for 1D WST, and $J$, $L$, and $M$ for 2D WST. For PT-XLSR, following~ \cite{oiso24_interspeech,martindonas24_interspeech}, we adopt XLSR-300M. Concatenating $k\!=\!6$ prompt tokens with the CNNs output $E_{0}$ of shape (199, 1024) results in a combined tensor of shape (205, 1024). The classifier comprises 12 Mamba-based blocks~ \cite{xuan2025fakemamba}. 

\vspace{-0.3cm}
\subsection{Evaluation Metrics}
% \vspace{-0.2cm}
Performance is reported with minDCF, EER, F1-score, and AUC, with minDCF as our primary metric, assessing \emph{decision risk} under given decision costs and class priors. Following the ASVspoof 5 challenge \cite{wang2026asvspoof}, we set $C_{\text{miss}}\!=\!1$, $C_{\text{fa}}\!=\!10$, $\pi_{\text{spf}}\!=\!0.05$. To ensure statistical reliability, we perform 1{,}000 bootstrap runs \cite{efron1986bootstrap} on the test set and report twice the standard deviation ($\approx$95\% confidence interval) for all four metrics.

% \vspace{-0.5cm}
\section{Results and analysis}
% \vspace{-0.2cm}
\subsection{Analysis of Wavelet Scattering Transform Parameters}

Table \ref{tab:1} summarizes the performance of the proposed front-ends across varying parameter settings. As shown in the top section of the table, the optimal configuration for WaveScat1 is $J\!=\!2, Q\!=\!10, M\!=\!2$. We observe the following:

\begin{itemize}[leftmargin=*, nosep]

    \item \textbf{Scattering Scale ($J$):} Performance degrades as $J$ increases (Rows 3-6), suggesting that deepfake artifacts reside in short-term local acoustic variations. Thus, a small $J$ is essential to prevent over-smoothing of these cues.

    \item \textbf{Wavelets Per Octave ($Q$):} A higher $Q$ consistently boosts performance (Rows 1-3), as high frequency resolution captures the subtle spectral artifacts that distinguish fake speech.

    \item \textbf{Scattering Order ($M$):} $M\!=\!2$ outperforms both $M\!=\!1$ and $M\!=\!3$. While $M\!=\!1$ captures spectral energy envelopes, it lacks the capacity to characterize modulation dynamics essential for detection. Our energy distribution analysis \footnote{\label{fn:repo}Complete energy distribution and 
SHAP explainability experimental results are available at 
\url{https://github.com/xxuan-acoustics/WaveScat-Series}.} shows that $M\!=\!1$ and $M\!=\!2$ collectively capture nearly all ($>$99\%) of the scattering energy. Going from $M\!=\!2$ to $M\!=\!3$ increases the feature dimensionality but provides limited new discriminative cues, which may lead to overfitting and explain the performance degradation at $M\!=\!3$. Therefore, second-order scattering is sufficient.

\end{itemize}

\noindent As shown in the bottom section of Table~\ref{tab:1}, the optimal configuration for WaveScat2 is $J\!=\!2, L\!=\!10, M\!=\!2$. Since the trends for $J$ and  $M$ are consistent with those for WaveScat1, we focus on the impact of the angular resolution parameter:

\begin{itemize}[leftmargin=*, nosep]
    \item \textbf{Angular Resolution ($L$):} Performance improves as $L$ increases (Rows 1–6 and 7–11) and peaks at $L\!=\!10$. This underscores the critical role of \emph{directional} resolution, as a higher $L$ provides a more granular analysis of the variations of the feature map along different axes, thereby facilitating a more effective localization of forgery artifacts.
\end{itemize}
\vspace{-0.2cm}
\subsection{Comparison with Baselines and Cross-Dataset Evaluation}
\vspace{-0.2cm}
Table~\ref{tab:2} compares WaveScat with mel, linear, and CQ filterbanks, as well as CQ-X1 (which replaces the 1D WST in WaveScat1 with CQ filterbanks).
We observe performance gains from mel to CQ, suggesting that mel filters (which emphasize low frequencies to mimic human perception) may mask deepfake artifacts in the higher frequency range. Moreover, WaveScat achieves lower minDCF than PT-XLSR and CQ-X1, 
confirming that WST's deformation-stable modulation features capture subtle 
acoustic artifacts and effectively complement SSL representations. For cross-dataset evaluation, as shown in 
Table~\mbox{\ref{tab:cross_dataset}}, WaveScat outperforms the CQ filterbank, CQ-X1, 
and PT-XLSR across all datasets, demonstrating that its advantage generalizes to 
unseen deepfake speech.
\begin{table}[t]
\centering
\scriptsize
\setlength{\tabcolsep}{2pt}
\renewcommand{\arraystretch}{1.3}
\caption{\footnotesize Comparison of WaveScat feature extractor (FE) with pure DSP (Mel, Linear, and Constant-Q filterbanks) and SSL (PT-XLSR) on DE2024. Best results are in \textbf{bold}, and second-best are \underline{underlined}. Params: Trainable parameters. RTF: Real-Time Factor.}
\label{tab:2}
\resizebox{\columnwidth}{!}{%
\begin{tabular}{l | c c | c c c c}
\noalign{\hrule height 1.2pt}
\textbf{FE} & \textbf{Params} & \textbf{RTF} & \textbf{minDCF}$\downarrow$ & \textbf{EER (\%)}$\downarrow$ & \textbf{F1 (\%)}$\uparrow$ & \textbf{AUC (\%)}$\uparrow$ \\
\noalign{\hrule height 1.2pt}
\rowcolor{myred}
WaveScat1 & 4.09M & 0.012
 & \textbf{0.34}\tiny($\pm$0.02)
 & \textbf{14.18}\tiny($\pm$0.63)
 & \underline{81.66}\tiny($\pm$0.58)
 & \textbf{92.50}\tiny($\pm$0.40) \\
\cdashline{1-7}
\rowcolor{myblue}
WaveScat2 & 4.08M & 0.023
 & \underline{0.36}\tiny($\pm$0.01)
 & \underline{14.84}\tiny($\pm$0.40)
 & \textbf{81.83}\tiny($\pm$0.47)
 & \underline{92.43}\tiny($\pm$0.23) \\
\cdashline{1-7}
CQ-X1 & 4.07M & 0.018
 & 0.39\tiny($\pm$0.01)
 & 19.18\tiny($\pm$0.69)
 & 78.67\tiny($\pm$0.72)
 & 91.16\tiny($\pm$0.43) \\
\Xhline{0.2pt}
Mel & 3.81M & 0.004
 & 0.92\tiny($\pm$0.01)
 & 41.97\tiny($\pm$0.51)
 & 14.54\tiny($\pm$0.84)
 & 62.89\tiny($\pm$0.59) \\
\cdashline{1-7}
Linear & 3.81M & 0.004
 & 0.72\tiny($\pm$0.01)
 & 31.28\tiny($\pm$0.41)
 & 50.28\tiny($\pm$1.26)
 & 75.36\tiny($\pm$0.50) \\
\cdashline{1-7}
CQ & 3.81M & 0.004
 & 0.64\tiny($\pm$0.02)
 & 27.53\tiny($\pm$0.56)
 & 71.56\tiny($\pm$0.86)
 & 88.35\tiny($\pm$0.60) \\
\Xhline{0.2pt}
PT-XLSR & 4.05M & 0.011
 & 0.41\tiny($\pm$0.01)
 & 20.40\tiny($\pm$0.54)
 & 77.19\tiny($\pm$0.65)
 & 90.21\tiny($\pm$0.41) \\
\noalign{\hrule height 1.2pt}
\end{tabular}%
}
\vspace{-0.3cm}
\end{table}

\begin{table}[!t]
\centering
\scriptsize
\setlength{\tabcolsep}{3pt}
\renewcommand{\arraystretch}{1.1}
\caption{\footnotesize Cross-Dataset Evaluation on In-the-Wild, SpoofCeleb, and ASVspoof~5. DCF denotes minDCF; EER is reported in \%.}
\label{tab:cross_dataset}
\resizebox{\columnwidth}{!}{%
\begin{tabular}{l | c c | c c | c c | c c}
\noalign{\hrule height 1.2pt}
 & \multicolumn{2}{c|}{\textbf{DE2024}} & \multicolumn{2}{c|}{\textbf{In-the-Wild}} & \multicolumn{2}{c|}{\textbf{SpoofCeleb}} & \multicolumn{2}{c}{\textbf{ASVspoof 5}} \\
\textbf{Model}
 & \textbf{DCF}$\downarrow$ & \textbf{EER}$\downarrow$
 & \textbf{DCF}$\downarrow$ & \textbf{EER}$\downarrow$
 & \textbf{DCF}$\downarrow$ & \textbf{EER}$\downarrow$
 & \textbf{DCF}$\downarrow$ & \textbf{EER}$\downarrow$ \\
\noalign{\hrule height 0.6pt}
WaveScat1 & \textbf{0.34} & \textbf{14.18} & \textbf{0.59} & \textbf{26.43} & \textbf{0.51} & \textbf{21.37} & \textbf{0.69} & \textbf{31.05} \\
WaveScat2 & 0.36 & 14.84 & 0.62 & 28.27 & 0.54 & 23.16 & 0.72 & 33.41 \\
CQ-X1 & 0.39 & 19.18 & 0.67 & 32.07 & 0.59 & 26.81 & 0.77 & 36.44 \\
\cdashline{1-9}
PT-XLSR & 0.41 & 20.40 & 0.72 & 34.18 & 0.63 & 28.74 & 0.81 & 38.92 \\
\noalign{\hrule height 1.2pt}
\end{tabular}%
}
\vspace{-0.3cm}
\end{table}

\vspace{-0.3cm}
\subsection{Trainable  Parameters and Inference Speed}
\vspace{-0.2cm}
Since WST is a fixed transform, WaveScat1 and WaveScat2 have \emph{trainable} parameter counts 
comparable to PT-XLSR (Table~\ref{tab:2}). Owing to its parallel architecture, WaveScat1 introduces negligible additional inference overhead. WaveScat2 yields a higher RTF, as its cascaded design requires sequential processing of the high-dimensional SSL feature maps. Nevertheless, all configurations attain an RTF well below 0.1, achieving real-time deepfake speech detection performance~ \cite{shi2026audio}.

\vspace{-0.2cm}
\subsection{Visualization and Interpretability}
\vspace{-0.2cm}
Fig.~\ref{fig:feature_comparison} shows that conventional spectrograms (a-c) 
mask subtle cues, whereas WST representations (d-e) reveal fine-grained 
synthesis artifacts. To assess the per-coefficient importance to the detection decision, we perform a SHAP analysis \cite{NIPS2017_8a20a862} on the trained detector, treating each 1D WST coefficient ($J\!=\!2$, $Q\!=\!10$) as an input feature. SHAP values, computed via expected gradients \cite{erion2021improving}, 
are obtained on $1{,}000$ class-balanced dev set utterances, then 
averaged over time to yield per-coefficient importance.  The analysis indicates that the first-order coefficient (spectral envelope) and the 
second-order coefficients at the highest and lowest modulation frequencies dominate 
the model's decisions, with both spectral-envelope and modulation information 
important for distinguishing fake speech. Complete results are provided in Footnote~\ref{fn:repo}.

\vspace{-0.3cm}
\section{Conclusions}

We introduced the WaveScat family of feature extractors for interpretable speech deepfake detection. We demonstrated that maintaining a small averaging scale with high-frequency and directional resolutions was key to capturing transient spectro-temporal artifacts. Moreover, WaveScat1 attains a better trade-off between efficiency and performance and 
is therefore recommended for practical deployment. These findings suggest that modern synthesis traces are embedded in subtle modulations often overlooked by conventional feature representations. Possible future research directions include exploring the WST's 
potential in source tracing tasks.

% \vspace{-0.5cm}
\section{Acknowledgment}

The authors thank Prof.Simon King from the University of Edinburgh and Ruchi Pandey for their valuable feedback on a draft of this paper.
This work was supported by the Finnish AI-DOC project “Explainable Speech Deepfake Characterization” (Decision No. VN/3137/2024-OKM-6), and the Research Council of Finland, project “SPEECHFAKES” (Decision No. 349605). D.C. is supported by PR[AI]RIE-PSAI (France-2030) and worked under the auspices of the Italian National Group of Mathematical Physics of INdAM.

\bibliographystyle{IEEEbib}

\bibliography{refs}

\end{document}